\documentclass[aps,prl,twocolumn,showpacs,showkeys,amsmath,amssymb]{revtex4}
\usepackage{amsfonts}
\usepackage{amsmath}
\usepackage{graphicx}
\usepackage{subfigure}
\usepackage{dcolumn}
\usepackage{bm}
\usepackage{booktabs}
\usepackage[utf8]{inputenc}

\usepackage{graphicx,graphics,dcolumn,booktabs,bm}
\usepackage{longtable,lscape}
\usepackage{txfonts}
\usepackage{overpic}
\usepackage{amssymb}
\usepackage{indentfirst}
\usepackage{epsfig}
\usepackage{feynmf}   
\usepackage{epstopdf}   
\usepackage{slashed}  
\usepackage{multirow}
\usepackage[colorlinks, citecolor=blue,anchorcolor=red,menucolor=red, linkcolor=red,filecolor=red,runcolor=red,urlcolor=blue,frenchlinks=red]{hyperref}

\usepackage{ulem}
\usepackage[usenames,dvipsnames]{color}

\makeatletter
\newcommand{\figcaption}{\def\@captype{figure}\caption}
\newcommand{\tabcaption}{\def\@captype{table}\caption}

\newcommand{\Rmnum}[1]{\expandafter\@slowromancap\romannumeral #1@}
\def\hlinewd#1{%
  \noalign{\ifnum0=`}\fi\hrule \@height #1 \futurelet
   \reserved@a\@xhline}
\makeatother
\def\dab{\int^{\alpha_{max}}_{\alpha_{min}}d\alpha\int^{\beta_{max}}_{\beta_{min}}d\beta}
\def\qq{\langle\bar qq\rangle}
\def\ss{\langle \bar ss\rangle}

\def\GGb{\langle g_s^2GG\rangle}
\def\qGqa{\langle\bar qg_s\sigma\cdot Gq\rangle}

\def\sGsa{\langle\bar sg_s\sigma\cdot Gs\rangle}

\def\f(s){[(\alpha+\beta)m^2-\alpha\beta s]}
\def\non{\\ \nonumber}

\begin{document}
%
%
\title{Decoding the $X(5568)$ as a fully open-flavor $su\bar b\bar d$ tetraquark state}

\author{Wei Chen$^1$}
\author{Hua-Xing Chen$^2$}
\email{hxchen@buaa.edu.cn}
\author{Xiang Liu$^{3,4}$}
\email{xiangliu@lzu.edu.cn}
\author{T. G. Steele$^1$}
\email{tom.steele@usask.ca}
\author{Shi-Lin Zhu$^{5,6,7}$}
\email{zhusl@pku.edu.cn}
\affiliation{
$^1$Department of Physics and Engineering Physics, University of Saskatchewan, Saskatoon, Saskatchewan, S7N 5E2, Canada
\\
$^2$School of Physics and Beijing Key Laboratory of Advanced Nuclear Materials and Physics, Beihang University, Beijing 100191, China
\\
$^3$School of Physical Science and Technology, Lanzhou University, Lanzhou 730000, China
\\
$^4$Research Center for Hadron and CSR Physics, Lanzhou University and Institute of Modern Physics of CAS, Lanzhou 730000, China
\\
$^5$School of Physics and State Key Laboratory of Nuclear Physics and Technology, Peking University, Beijing 100871, China
\\
$^6$Collaborative Innovation Center of Quantum Matter, Beijing 100871, China
\\
$^7$Center of High Energy Physics, Peking University, Beijing 100871, China
}

\begin{abstract}
We investigate the recent evidence for a charged $X(5568)$ meson as an exotic open-flavor tetraquark
state $su\bar b\bar d$ with $J^P=0^+/1^+$ in the framework of QCD sum rules. We use the color
antisymmetric $[\mathbf{\bar 3_c}]_{su} \otimes [\mathbf{3_c}]_{\bar{b}\bar d}$ tetraquark currents
in both scalar and axial-vector channels to perform evaluations and numerical analyses. Our results
imply that the $X(5568)$ can be interpreted as both the scalar $su\bar b\bar d$ tetraquark state and
the axial-vector one, which are in good agreement with the experimental measurement. We also discuss
the possible decay patterns of the $X(5568)$ and suggest to search for its neutral partner in the radiative
decay into $B_s^0 \gamma$ and $B_s^* \gamma$, which can be used to determine its spin-parity
quantum numbers. Moreover, we predict its charmed partner state around $2.55$ GeV with the quark
content $su\bar c\bar d$ and $J^P=0^+/1^+$.
\end{abstract}


 \maketitle
%
%

{\it{Introduction}}.---As a long-standing puzzle, understanding the nonperturbative QCD behavior quantitatively is one of the most important and intriguing research topics of the hadron physics. Thoroughly figuring out the map of hadron spectrum is a key step to achieve this goal. The possible hadron configurations not only include conventional mesons and baryons, but also contain exotic states like glueball, hybrid, and multiquark states, etc. However, exotic states have not yet been well established, which is the reason why experimentalists and theorists have already paid many and are still paying more attentions to them. With significant experimental progress in the past decade, more and more charmonium-like and bottomonium-like states (also named as $XYZ$ particles)~\cite{Agashe:2014kda}, and the hidden-charm pentaquarks $P_c(4380)$ and $P_c(4450)$ \cite{Aaij:2015tga} were observed, which provide good chance for identifying exotic states (see Refs. \cite{Liu:2013waa,Chen:2016qju} for review). It is obvious that this story still continues with recent evidence
for a $X(5568)$ state~\cite{D0:2016mwd}.

The $X(5568)$ is a narrow structure seen by the D\O~~Collaboration \cite{D0:2016mwd}, which appears in the $B_s^0\pi^\pm$ invariant mass spectrum with 5.1$\sigma$ significance. Its measured mass and width are $M=5567.8\pm2.9(\rm stat)^{+0.9}_{-1.9}(\rm syst)$ MeV and $\Gamma=21.9\pm6.4(\rm stat)^{+5.0}_{-2.5}(\rm syst)$ MeV, respectively. Its decay final state $B_s^0\pi^\pm$ requires the valence quark component of the $X(5568)$ to be $su\bar{b}\bar{d}$ (or $sd\bar{b}\bar{u}$). Hence, the reported $X(5568)$ state, if exist,
cannot be categorized into the conventional meson family, and is a good candidate of exotic tetraquark state with
valence quarks of four different flavors.

With the mass of $D_{sj}^+(2632)$~\cite{Evdokimov:2004iy} as input, the
scalar isoscalar
$b {\bar s} q {\bar q}$ tetraquark mass was estimated to be around 5832 MeV
in Ref.~\cite{Liu:2004kd},
while the mass of the isovector $b {\bar s} u {\bar d}$ tetraquark state can
be similarly estimated to be around 5700 MeV.
In Ref.~\cite{Kolomeitsev:2003ac}, the heavy-light meson resonances with $J^P=0^+$ and $J^P=1^+$
and similar flavor configurations as the $X(5568)$ were studied in terms of the non-linear
chiral SU(3) Lagrangian and their masses were estimated
around 5750-5790 MeV, which is 200 MeV higher than the mass of the $X(5568)$.
The $X(5568)$ was studied as a $su\bar b\bar d$ tetraquark state with $J^P=0^+$
in Ref. \cite{Agaev:2016mjb}.
Inspired by the charm-strange state $D_{s0}^*(2317)$~\cite{Aubert:2003fg},
various exotic pictures were proposed such as the $cq\bar s\bar q$ tetraquark state~\cite{Maiani:2004vq,Terasaki:2003qa,Wang:2006uba,Vijande:2006hj}.

In this work, we investigate the $X(5568)$ as a fully open-flavor
$su\bar b\bar d$ tetraquark state with $J^P=0^+/1^+$ in the framework of QCD sum rules.
Our results suggest that the interpretations of the $X(5568)$ as the scalar $su\bar b\bar d$ tetraquark state
and the axial-vector one are both possible. To differentiate them and determine its spin-parity quantum numbers, we further investigate its possible decay patterns, from which
we propose to observe the radiative decay of the neutral partner of the $X(5568)$
into $B_s^0 \gamma$ and $B_s^* \gamma$.
The charmed partner state of the $X(5568)$, with the fully open-flavor
quark content $su\bar c\bar d$, can be searched
for in many current experiments.
We predict the mass of this charmed-partner to be around 2.55 GeV
for both the cases of $J^P=0^+$ and $1^+$.

{\it{Interpretation of the $X(5568)$ state}}.--- We shall first construct the diquark-antidiquark
type of tetraquark interpolating currents with quark content $su\bar{b}\bar{d}$. There are five independent
diquark fields: $q^T_aCq_b$, $q^T_a C\gamma_5q_b$, $q^T_aC\gamma_\mu q_b$,
$q^T_aC\gamma_\mu\gamma_5q_b$ and $q^T_a C\sigma_{\mu\nu}q_b$, where $a, b$ are color indices. In general, one can use all
these diquarks and the corresponding antidiquarks to compose tetraquark operators.
However, the $P$-wave diquarks $q^T_aCq_b$, $q^T_aC\gamma_\mu\gamma_5q_b$ and
$q^T_a C\sigma_{\mu\nu}q_b$ are not favored configurations \cite{Jaffe:2004ph}. In QCD sum rules,
they lead to unstable sum rules and thus unreliable mass predictions \cite{Chen:2010ze,Kleiv:2013dta}.
Following our previous works in Refs. \cite{Chen:2013aba,Du:2012wp},
we use only the $S$-wave diquark fields $q_a^TC\gamma_5q_b$ and
$q_a^TC\gamma_{\mu}q_b$ to compose the tetraquark currents coupling to the lowest lying
hadron states with $J^P=0^+/1^+$.
Considering the Lorentz and color structures, we finally obtain the interpolating currents with
$J^P=0^+$
\begin{equation}
\begin{split}
J_1&=s^T_aC\gamma_5u_b(\bar{b}_a\gamma_5C\bar{d}^T_b+\bar{b}_b\gamma_5C\bar{d}^T_a)\, ,\\
J_2&=s^T_aC\gamma_\mu u_b(\bar{b}_a\gamma^\mu
C\bar{d}^T_b+\bar{b}_b\gamma^\mu C\bar{d}^T_a)\, ,
\\
J_3&=s^T_aC\gamma_5u_b(\bar{b}_a\gamma_5C\bar{d}^T_b-\bar{b}_b\gamma_5C\bar{d}^T_a)\, ,\\
J_4&=s^T_aC\gamma_\mu u_b(\bar{b}_a\gamma^\mu
C\bar{d}^T_b-\bar{b}_b\gamma^\mu C\bar{d}^T_a)\, , \label{currents0+}
\end{split}
\end{equation}
in which $J_1$ and $J_2$ belong to the symmetric color structure
$[\mathbf{6_c}]_{su} \otimes [\mathbf{ \bar 6_c}]_{\bar{b}\bar d}$,
and $J_3$ and $J_4$ belong to the antisymmetric color structure
$[\mathbf{\bar 3_c}]_{su} \otimes [\mathbf{3_c}]_{\bar{b}\bar d}$.
The interpolating currents with $J^P=1^+$ are
\begin{equation}
\begin{split}
J_{1\mu}&=s^T_aC\gamma_5u_b(\bar{b}_a\gamma_\mu C\bar{d}_b^T+\bar{b}_b\gamma_\mu C\bar{d}_a^T)\, ,\\
J_{2\mu}&=s^T_aC\gamma_\mu u_b(\bar{b}_a\gamma_5C\bar{d}_b^T+\bar{b}_b\gamma_5C\bar{d}^T_a)\, ,\\
J_{3\mu}&=s^T_aC\gamma_5u_b(\bar{b}_a\gamma_\mu C\bar{d}_b^T-\bar{b}_b\gamma_\mu C\bar{d}_a^T)\, ,\\
J_{4\mu}&=s^T_aC\gamma_\mu u_b(\bar{b}_a\gamma_5C\bar{d}_b^T-\bar{b}_b\gamma_5C\bar{d}^T_a)\, ,
\label{currents1+}
\end{split}
\end{equation}
in which $J_{1\mu}$ and $J_{2\mu}$ belong to the symmetric color structure
$[\mathbf{6_c}]_{su} \otimes [\mathbf{ \bar 6_c}]_{\bar{b}\bar d}$,
and $J_{3\mu}$ and $J_{4\mu}$ belong to the antisymmetric color structure
$[\mathbf{\bar 3_c}]_{su} \otimes [\mathbf{3_c}]_{\bar{b}\bar d}$.
We note that these currents can be related to those $cq\bar b\bar q$ currents in Ref. \cite{Chen:2013aba}
by simply replacing the charm quark $c$ by the strange quark $s$.

In the following, we use these currents to study the fully open-flavor $su\bar{b}\bar{d}$ tetraquark states
in the framework of QCD sum rules.
Arising from QCD itself, QCD sum rule techniques provide a model-independent method to study nonperturbative
problems in strong interaction physics. In the past several decades, QCD sum-rules have been used to study a vast number of hadronic properties for conventional mesons and baryons as reviewed in Refs. \cite{Shifman:1978bx,Reinders:1984sr,colangelo,Nielsen:2009uh,Narison:2002pw}.
Recently, QCD sum-rules have been also extended to studies of exotic hadron states and the results are encouraging~\cite{Nielsen:2009uh}.
However, the accuracy of the method is limited by the truncation of the OPE series and the complicated
structure of the hadronic dispersion integrals.

We start from the two-point correlation functions
\begin{align}
\Pi(p^{2})&= i\int d^4xe^{ip\cdot
x}\langle0|T[J(x)J^{\dag}(0)]|0\rangle\, , \label{equ:Pi1}
\non \Pi_{\mu\nu}(p^{2})&= i\int d^4xe^{ip\cdot x}\langle0|T[J_{\mu}(x)J_{\nu}^{\dag}(0)]|0\rangle
\\ &=\left(\frac{p_{\mu}p_{\nu}}{p^2}-g_{\mu\nu}\right)\Pi_1(p^2)+\frac{p_{\mu}p_{\nu}}{p^2}\Pi_0(p^2)\, , \label{equ:Pi2}
\end{align}
for scalar and axial-vector tetraquark systems, respectively. The imaginary parts of the invariant
functions $\Pi_1(p^2)$ and $\Pi_0(p^2)$ receive contributions from the pure spin-1 and spin-0
intermediate states, respectively. In this paper we shall use $\Pi(p^{2})$ and $\Pi_1(p^2)$ to investigate
the scalar and axial-vector channels, respectively.

One can build QCD sum rules on the hypothesis that the two-point correlation function can be evaluated
at the quark-gluonic level, which is then equated to that obtained at the hadronic level. The correlation
function at the hadronic level can be described in the form of the dispersion relation
\begin{align}
\Pi(p^2)=\frac{(p^2)^N}{\pi}\int_{<}^{\infty}\frac{\mbox{Im}\Pi(s)}{s^N(s-p^2-i\epsilon)}ds
+\sum_{n=0}^{N-1}b_n(p^2)^n\, ,
\label{dispersionrelation}
\end{align}
where $b_n$ are subtraction constants. These unknown constants can be removed later
by performing the Borel transform. The imaginary part of the correlation function
is defined as the spectral density $\rho(s)\equiv{\mbox{Im}\Pi(s)}/{\pi}$, which can be written
as a sum over $\delta$ functions by inserting intermediate hadronic states
\begin{align}
\nonumber
\rho(s)&=\sum_n\delta(s-m_n^2)\langle0|J|n\rangle\langle
n|J^{\dagger}|0\rangle+\text{continuum}
\\&=f_X^2\delta(s-m_X^2)+ \mbox{continuum} \, , \label{Imaginary}
\end{align}
in which a narrow resonance approximation is adopted in the second step.
The interpolating current $J(x)$ can couple to all intermediate hadrons
$|n\rangle$ with the same quantum numbers.
As usual we only investigate the lowest lying resonance $|X\rangle$,
and $m_X$ and $f_X$ are its mass and coupling constant, respectively.
The coupling constants are defined as
\begin{eqnarray}
\langle0|J|X\rangle&=&f_X\, , \label{coupling parameter1}
\\
\langle0|J_{\mu}|X\rangle&=&f_X\epsilon_{\mu}\, , \label{coupling
parameter2}
\end{eqnarray}
for the scalar and axial-vector interpolating currents $J(x)$ and $J_\mu(x)$,
respectively. $\epsilon_{\mu}$ is a polarization vector ($\epsilon\cdot p=0$).

At the quark-gluonic level, the correlation functions $\Pi(p^{2})$ and $\Pi_1(p^2)$ can be evaluated
using the method of the operator product expansion (OPE).
In this paper, we shall use the
interpolating currents $J_3(x)$ with $J^P=0^+$ and $J_{3\mu}(x)$ with
$J^P=1^+$ to investigate the
$X(5568)$ as a $su\bar b\bar d$ tetraquark state.
The spectral density for the current $J_3(x)$ with $J^P=0^+$ is
{\allowdisplaybreaks
\begin{eqnarray}
\rho_0(s)&=&\frac{1}{512\pi^6}\dab (1-\alpha-\beta)^2
\non&&
\frac{(m_b^2\beta+m_s^2\alpha-3\alpha\beta
s)(m_b^2\beta+m_s^2\alpha-\alpha\beta
s)^3}{\alpha^3\beta^3}\, ,
\non
\rho_3(s)&=&-\frac{\qq}{16\pi^4}\dab\frac{(m_b^2\beta+m_s^2\alpha-\alpha\beta
s)}{\alpha\beta}
\non&&\left(\frac{m_b}{\alpha}+\frac{m_s}{\beta}\right)(1-\alpha-\beta)(m_b^2\beta
+m_s^2\alpha-2\alpha\beta s)
\non &&
+\frac{m_s\ss}{64\pi^4}\int_0^{1-m_b^2/s}d\alpha\frac{\alpha^2}{1-\alpha}
\non
&&\left[m_b^2-s(1-\alpha)\right]\left[m_b^2-2s(1-\alpha)\right]\, ,
\non
\rho_4(s)&=&\frac{\GGb}{1024\pi^6}\dab \Bigg\{\frac{(1-\alpha-\beta)^2}{3}
\non&&
(2m_b^2\beta+2m_s^2\alpha-3\alpha\beta
s)\left(\frac{m_b^2}{\alpha^3}+\frac{m_s^2}{\beta^3}\right)
\non &&
+\frac{(1-\alpha-\beta)(m_b^2\beta+m_s^2\alpha-2\alpha\beta s)}{\alpha\beta}
\non&& \left(\frac{1}{\alpha}+\frac{1}{\beta}\right)
(m_b^2\beta+m_s^2\alpha-\alpha\beta s)\Bigg\}\, ,
\non
\rho_5(s)&=&\frac{\qGqa}{64\pi^4}\dab
\non &&
(2m_b^2\beta+2m_s^2\alpha-3\alpha\beta s)
\non &&
\Bigg[\left(\frac{m_b}{\alpha}+\frac{m_s}{\beta}\right)
-\left(\frac{m_b}{\alpha^2}+\frac{m_s}{\beta^2}\right)
(1-\alpha-\beta)\Bigg]\, ,
\non
\non
\rho_6(s)&=&\frac{m_bm_s\qq^2}{12\pi^2}\left[\left(1+\frac{m_b^2-m_s^2}{s}\right)^2-\frac{4m_b^2}{s}\right]^{1/2}-
\non &&
\frac{\ss\qq}{24\pi^2}\int_0^{1-\frac{m_b^2}{s}}d\alpha
\left[4m_b^2\alpha-6s\alpha(1-\alpha)+m_bm_s\right]\, ,
\non
\rho_8(s)&=&\frac{\ss\qGqa+\sGsa\qq}{48\pi^2}
\non &&
+\frac{\qq\qGqa}{24\pi^2}\int_0^1d\alpha
\non&&
\Bigg[\frac{m_bm_s^3}{\alpha^2}\delta'\left(s-\frac{m_b^2\alpha+m_s^2(1-\alpha)}{\alpha(1-\alpha)}\right)
\non &&
-\frac{m_bm_s}{2\alpha(1-\alpha)}\delta\left(s-\frac{m_b^2\alpha+m_s^2(1-\alpha)}{\alpha(1-\alpha)}\right)\Bigg]\, .
\label{spectraldensity0+}
\end{eqnarray}
The spectral density for the current $J_{3\mu}(x)$ with $J^P=1^+$ is
{\allowdisplaybreaks
\begin{eqnarray}
\rho_0(s)&=&\frac{1}{1024\pi^6}\dab(1-\alpha-\beta)^2
\non&&\frac{(m_b^2\beta+m_s^2\alpha-5\alpha\beta
s)(m_b^2\beta+m_s^2\alpha-\alpha\beta
s)^3}{\alpha^3\beta^3}\, ,
\non
\rho_3(s)&=&\frac{\qq}{16\pi^4}\dab(m_b^2\beta+m_s^2\alpha-\alpha\beta
s)
\non&&
(1-\alpha-\beta)\Bigg[\frac{m_s(m_b^2\beta+m_s^2\alpha-\alpha\beta
s)}{2\alpha\beta^2}
\non&&-\left(\frac{m_b}{\alpha}+\frac{m_s}{\beta}\right)\frac{(m_b^2\beta+m_s^2\alpha-2\alpha\beta
s)}{\alpha\beta}\Bigg]
\non&&
+\frac{m_s\ss}{128\pi^4}\int_0^{1-m_b^2/s}d\alpha
\non
&&\frac{\alpha^2\left[m_b^2-s(1-\alpha)\right]\left[m_b^2-3s(1-\alpha)\right]}{1-\alpha}\, , \non
\rho_4(s)&=&\frac{\GGb}{3072\pi^6}\dab\Bigg[(1-\alpha-\beta)^2
\non &&
(m_b^2\beta+m_s^2\alpha-2\alpha\beta
s)\left(\frac{m_b^2}{\alpha^3}+\frac{m_s^2}{\beta^3}\right)
\non&&-\frac{(1-\alpha-\beta)(m_b^2\beta+m_s^2\alpha-\alpha\beta
s)}{2\alpha\beta}
\non &&
\Bigg(\frac{3m_b^2\beta+3m_s^2\alpha-5\alpha\beta
s}{\alpha}
\non&&
-\frac{3m_b^2\beta+3m_s^2\alpha-9\alpha\beta
s}{\beta}\Bigg)\Bigg]\, ,
\non
\rho_5(s)&=&\frac{\qGqa}{64\pi^4}\dab
\non&&
\Bigg[\frac{m_b(2m_b^2\beta+2m_s^2\alpha-3\alpha\beta
s)}{\alpha}
\non&&-\frac{m_s(1-\alpha-2\beta)(m_b^2\beta+m_s^2\alpha-2\alpha\beta
s)}{\beta^2}\Bigg]\, ,
\non
\rho_6(s)&=&\frac{m_bm_s\qq^2}{12\pi^2}\left[\left(1+\frac{m_b^2-m_s^2}{s}\right)^2-\frac{4m_b^2}{s}\right]^{1/2}-
\non &&
\frac{\ss\qq}{24\pi^2}\int_0^{1-\frac{m_b^2}{s}}d\alpha
\left[2m_b^2\alpha-4s\alpha(1-\alpha)+m_bm_s\right]\, ,
\non
\rho_8(s)&=&\frac{\ss\qGqa+\sGsa\qq}{96\pi^2}\left(1+\frac{m_b^2}{s^2}\right)
\non&&
+\frac{\qq\qGqa}{24\pi^2}\int_0^1d\alpha
\non&&
\Bigg[\frac{m_bm_s^3}{\alpha^2}\delta'\left(s-\frac{m_b^2\alpha+m_s^2(1-\alpha)}{\alpha(1-\alpha)}\right)
\non&&
-\frac{m_bm_s}{2\alpha}\delta\left(s-\frac{m_b^2\alpha+m_s^2(1-\alpha)}{\alpha(1-\alpha)}\right)\Bigg]\, ,
\end{eqnarray}
}
where the subscript $i$ of $\rho_i(s)$ denotes the dimension of the condensate and $\rho_0(s)$ is the perturbative
term. The integral limitations are
\begin{eqnarray*}
\alpha_{min} &=& \frac{1}{2}\left\{1+\frac{m_b^2-m_s^2}{s}-\left[\left(1+\frac{m_b^2-m_s^2}{s}\right)^2-\frac{4m_b^2}{s}\right]^{1/2}\right\} \, ,
\\ \alpha_{max} &=& \frac{1}{2}\left\{1+\frac{m_b^2-m_s^2}{s}+\left[\left(1+\frac{m_b^2-m_s^2}{s}\right)^2-\frac{4m_b^2}{s}\right]^{1/2}\right\} \, ,
\\ \beta_{min} &=& \frac{\alpha m_s^2}{\alpha s-m_b^2} \, ,
\\ \beta_{max} &=& 1-\alpha \, .
\end{eqnarray*}

We adopt the following parameter values to perform QCD sum rule analysis \cite{Reinders:1984sr,Agashe:2014kda,Narison:2011rn,Narison:2010cg,Kuhn:2007vp,Chen:2015fwa}
in the chiral limit $(m_u=m_d=0)$:
\begin{equation}
\begin{split}
& m_s(2\,\text{GeV})=(95\pm 5)\text{ MeV} \, ,
\\ &
m_c(m_c)=\overline m_c=(1.275\pm 0.025) \mbox{ GeV}   \, ,
\\ &
m_b(m_b)=\overline m_b=(4.18\pm 0.03) \mbox{ GeV}   \, ,
\\&
\qq=-(0.24\pm0.01)^3\text{ GeV}^3 \, ,
\\&
\ss=(0.8\pm0.1)\qq \, ,
\\ &
\qGqa=-M_0^2\qq \, ,
\\ &
\sGsa=-M_0^2\ss \, ,
\\ &
M_0^2=(0.8\pm0.2)\text{ GeV}^2 \, ,
\\ &
\GGb=(0.48\pm0.14) \text{ GeV}^4
\, , \label{parameters}
\end{split}
\end{equation}
in which $\overline m_c$ and $\overline m_b$ are the ``running masses" of the heavy quarks in the $\overline{\rm MS}$ scheme.
There is an additional minus sign in mixed condensates due to the different definition of the
coupling constant $g_s$ compared to that in Ref.~\cite{Reinders:1984sr}.

In the following, we use $J_3(x)$ and $J_{3\mu}(x)$ to perform
numerical analyses.
To establish the QCD sum rules, the Borel transform is usually performed on the correlation functions to pick
out the lowest lying state and remove the unknown subtraction constants $b_n$  in Eq. \eqref{dispersionrelation}.
Comparing  $\Pi(p^2)$ at both phenomenological and OPE sides, one can obtain the following QCD sum
rules via the quark-hadron duality
\begin{eqnarray}
\mathcal{L}_{k}\left(s_0,
M_B^2\right)=f_X^2m_X^{2k}e^{-m_X^2/M_B^2}=\int_{<}^{s_0}dse^{-s/M_B^2}\rho(s)s^k\, ,
\label{sumrule}
\end{eqnarray}
where $s_0$ and $M_B$ are the continuum threshold and Borel parameter.
The hadron mass is then extracted as
\begin{eqnarray}
m_X\left(s_0, M_B^2\right)=\sqrt{\frac{\mathcal{L}_{1}\left(s_0,
M_B^2\right)}{\mathcal{L}_{0}\left(s_0, M_B^2\right)}}\, . \label{mass}
\end{eqnarray}
One notes that the hadron mass in Eq. \eqref{mass} is a function of $s_0$ and $M_B$,
which are two essential parameters in our following analysis. The QCD sum rule prediction
of the hadron mass is only significant and reliable in suitable regions of the parameter space
$(s_0, M_B^2)$.

We use two criteria to fix the Borel parameter: a) the lower bound on $M_B^2$ can be determined by the constraint of the
OPE convergence, requiring the contribution of the dominant non-perturbative terms
(quark condensates here) to be less than one third of perturbative contribution;
b) the limitation of the pole contribution (PC) gives the upper bound on $M_B^2$
\begin{eqnarray}
\text{PC}(s_0, M_B^2)=\frac{\mathcal{L}_{0}\left(s_0,
M_B^2\right)}{\mathcal{L}_{0}\left(\infty, M_B^2\right)}\, .
\label{PC}
\end{eqnarray}
However, the extracted hadron mass $m_X$ should not depend on the
unphysical parameter $M_B$, which results in the stability criterion: we choose the
value of the continuum threshold $s_0$ to minimize the dependence of $m_X$ with
respect to the Borel mass $M_B$.
Altogether, we will obtain a Borel window $M_{min}^2\leq M_B^2\leq M_{max}^2$
for a definite value of $s_0$.

For the interpolating current $J_3(x)$ with $J^P=0^+$, we show the variation of the
extracted mass $m_X$ with respect to $s_0$ in the left panel of Fig. \ref{fig:mass0+}.
We find that the $M_B$ dependence of the hadron mass becomes very weak for
$32$ GeV$^2$ $\leq s_0\leq 36$ GeV$^2$, which is thus a reasonable working region.
Accordingly, we fix $s_0=34$ GeV$^2$, which is then used to fix the Borel window
$6.0$ GeV$^2$ $\leq M_B^2\leq 7.4$ GeV$^2$ using the two criteria discussed above.

Within the above parameter regions, we show the variation of the hadron mass $m_X$
with respect to the Borel parameter in the right panel of Fig. \ref{fig:mass0+}. We find that
the mass sum rules are very stable in the Borel window $6.0$ GeV$^2$ $\leq
M_B^2\leq 7.4$ GeV$^2$, and the hadron mass is extracted as
\begin{eqnarray}
m_{X,\, 0^+}=5.58\pm0.14 \text{ GeV}\, , \label{mass0+}
\end{eqnarray}
where the error comes from the uncertainties of the bottom quark mass $m_b$, the condensates
 $\qq, \ss, \GGb, \qGqa, \sGsa$, the parameter $M_0^2$ in Eq. \eqref{parameters} and the continuum
 threshold $s_0$. This mass value is in good agreement with the
observed mass of the $X(5568)$ state in Ref.~\cite{D0:2016mwd}, and implies
a possible $J^P = 0^+$ $su\bar b\bar d$ tetraquark interpretation for this exotic state.

\begin{figure}
\begin{tabular}{lr}
\scalebox{0.35}{\includegraphics{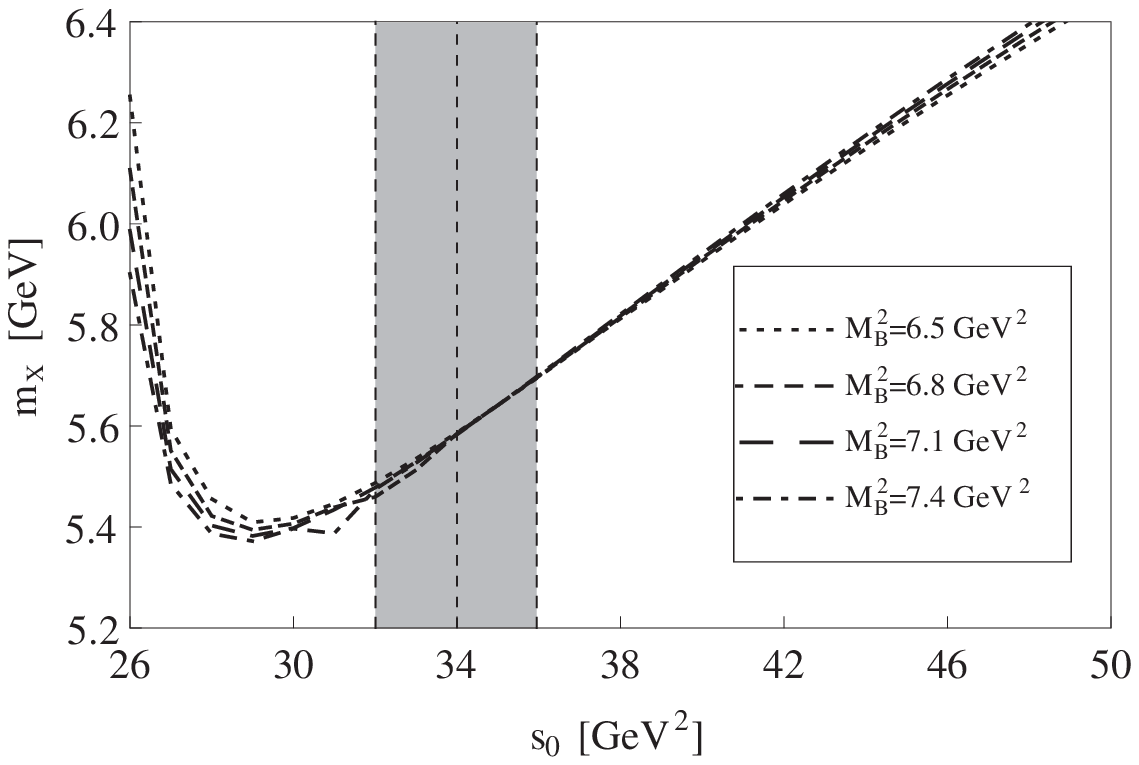}}&
\scalebox{0.35}{\includegraphics{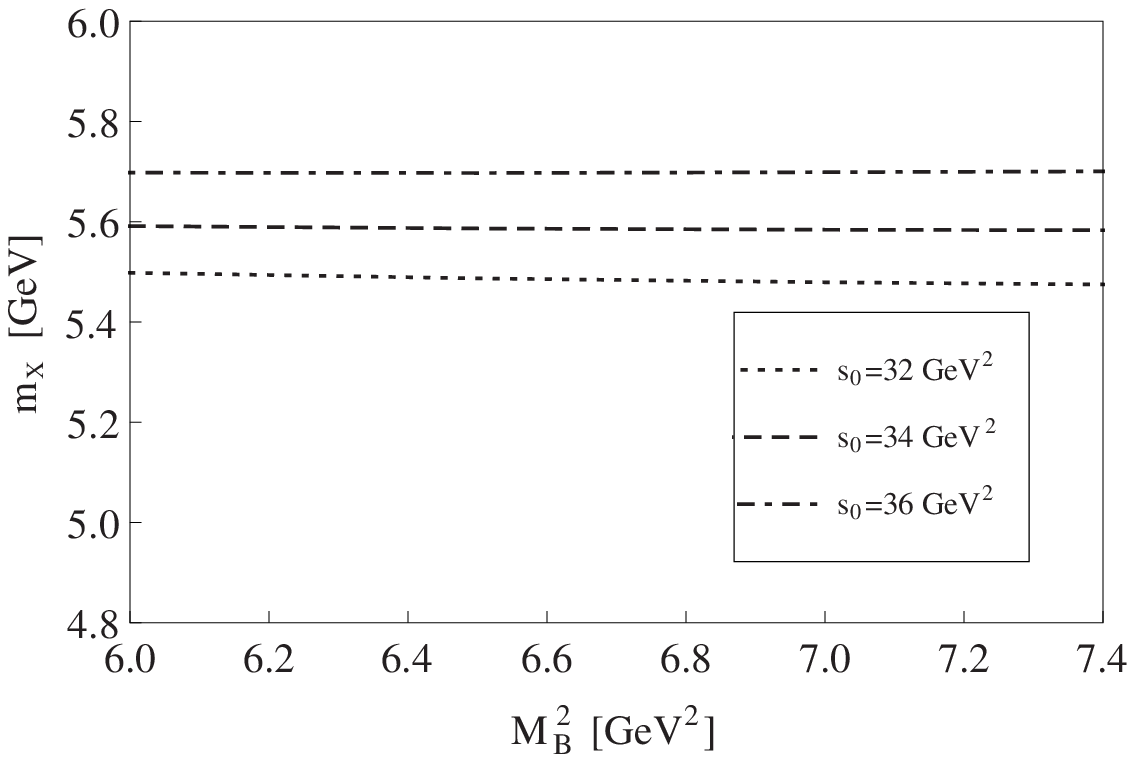}}
\end{tabular}
\caption{Variations of the extracted hadron mass with respect to the continuum threshold $s_0$ (left)
and the Borel parameter $M_B^2$ (right) for $J_{3}(x)$ with $J^{P}=0^{+}$.}
\label{fig:mass0+}
\end{figure}

For the axial-vector current $J_{3\mu}(x)$ with $J^P=1^+$, the OPE series has a similar
behavior to that of the scalar current $J_{3}(x)$. After performing the QCD sum rule analyses,
we find the working regions for the continuum threshold $32$ GeV$^2$ $\leq s_0\leq 36$
GeV$^2$ and Borel parameter $6.3$ GeV$^2$ $\leq M_B^2 \leq 7.4$ GeV$^2$.
We show the hadron mass $m_X$ as a function of $s_0$ and
$M_B$ in the above working regions in Fig. \ref{fig:mass1+}. These mass curves have
similar behavior to those in Fig. \ref{fig:mass0+}, and
the hadron mass is extracted as
\begin{eqnarray}
m_{X,\, 1^+}=5.59\pm0.15 \text{ GeV}\, . \label{mass1+}
\end{eqnarray}
This mass value is also in good agreement with the
measured mass of the $X(5568)$ state in Ref.~\cite{D0:2016mwd}, and supports
the axial-vector $su\bar b\bar d$ tetraquark interpretation for this meson.

For the other interpolating currents $J_1(x), J_2(x), J_4(x)$ with $J^P=0^+$ and $J_{1\mu}(x),
J_{2\mu}(x), J_{4\mu}(x)$ with $J^P=1^+$ in Eqs.~\eqref{currents0+}-\eqref{currents1+}, we
have also performed QCD sum rule analyses. We
shall detailly discuss them in our future studies, but just note that
the hadron masses extracted by using these currents are higher than the mass of the
$X(5568)$. Especially, the scalar currents $J_1(x), J_2(x)$ and axial-vector currents $J_{1\mu}(x),
J_{2\mu}(x)$ lead to hadron masses above 6 GeV, which may be due to that the color
structures of these tetraquark currents are symmetric
$[\mathbf{6_c}]_{su} \otimes [\mathbf{ \bar 6_c}]_{\bar{b}\bar d}$,
while those of $J_{3}(x)$ and $J_{3\mu}(x)$ used above are antisymmetric $[\mathbf{\bar 3_c}]_{su} \otimes
[\mathbf{3_c}]_{\bar{b}\bar d}$,
and the former color symmetric
diquarks and antidiquarks probably have larger masses than the latter
antisymmetric ones.

\begin{figure}
\begin{tabular}{lr}
\scalebox{0.35}{\includegraphics{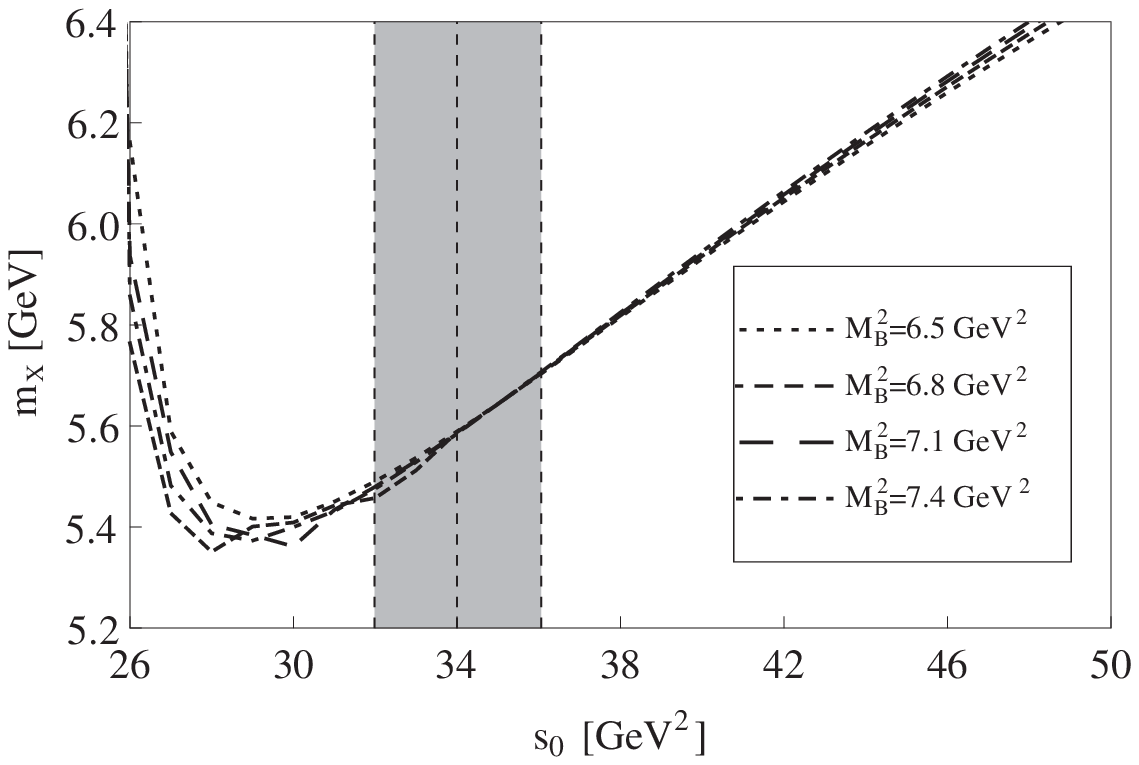}}&
\scalebox{0.35}{\includegraphics{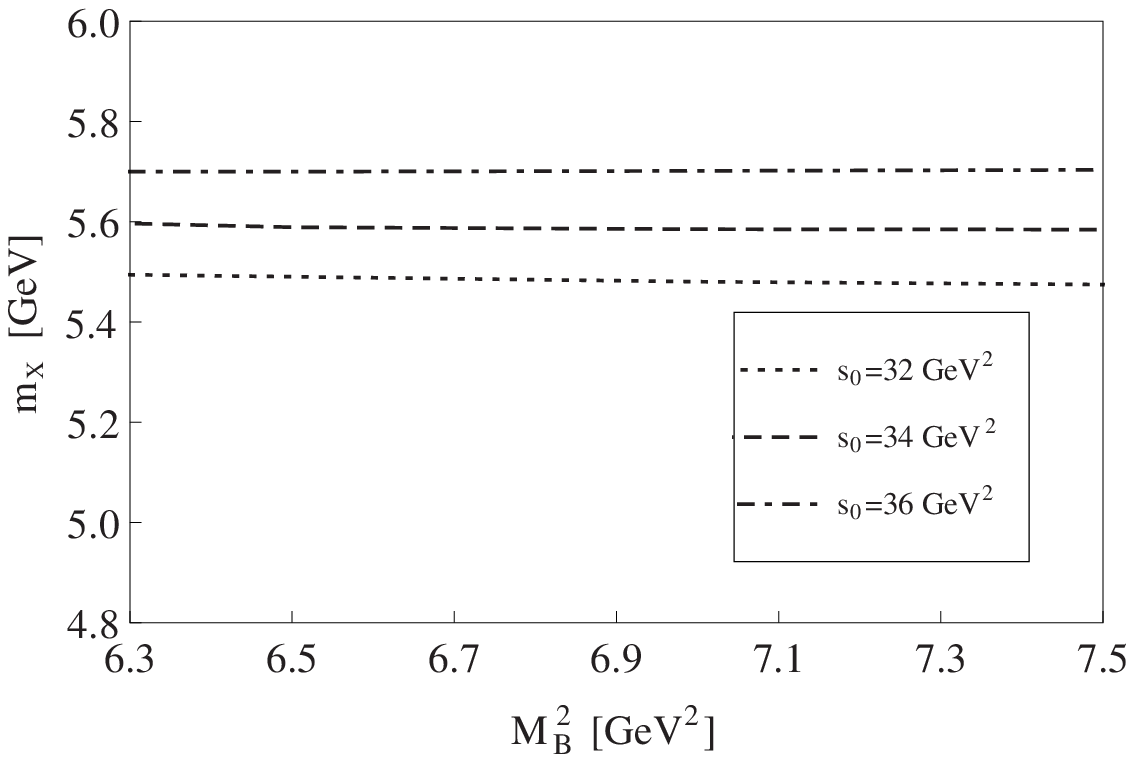}}
\end{tabular}
\caption{Variations of the extracted hadron mass with respect to the continuum threshold $s_0$ (left)
and the Borel parameter $M_B^2$ (right) for $J_{3\mu}(x)$ with $J^{P}=1^{+}$.}
\label{fig:mass1+}
\end{figure}

{\it{Decay patterns of the $X(5568)$ state}}.---As discussed in Refs.~\cite{Chen:2006hy,Chen:2015fwa}, the diquark-antidiquark tetraquark currents
can be transformed to be the mesonic-mesonic currents through
the Fierz transformation, from which we can study the decay patterns of the $X(5568)$:
\begin{enumerate}

\item The current $J_3(x)$ can be transformed into
\begin{eqnarray}
\nonumber J_3(x) &\rightarrow& \bar{b}_a s_a \bar{d}_b u_b \oplus \bar{b}_a \gamma_5 s_a \bar{d}_b \gamma_5 u_b \oplus \bar{b}_a \sigma_{\mu\nu} s_a \bar{d}_b \sigma^{\mu\nu} u_b
\\ \nonumber &\oplus& \bar{b}_a \gamma_\mu s_a \bar{d}_b \gamma^\mu u_b \oplus \bar{b}_a \gamma_\mu \gamma_5 s_a \bar{d}_b \gamma^\mu \gamma_5 u_b
\\ &\oplus& \{ s \leftrightarrow u \} \, .
\end{eqnarray}
This suggests that if the $X(5568)$ is a $su\bar b\bar d$ tetraquark state of $J^P=0^+$,
its kinematically allowed decay channel would be only the $S$-wave $B_s^0 \pi^+$,
which is just its observed channel~\cite{D0:2016mwd}.
Besides this, its neutral partner may also decay into $B_s^* \gamma$ through $\bar q \gamma_\mu q \rightarrow \gamma$.

\item The current $J_{3\mu}(x)$ can be transformed into
\begin{eqnarray}
\nonumber J_{3\mu}(x) &\rightarrow& \bar{b}_a s_a \bar{d}_b \gamma_\mu \gamma_5 u_b \oplus \bar{b}_a \gamma_5 s_a \bar{d}_b \gamma_\mu u_b
\\ \nonumber &\oplus& \bar{b}_a \gamma^\nu \gamma_5 s_a \bar{d}_b \sigma_{\mu\nu} u_b \oplus \bar{b}_a \gamma^\nu s_a \bar{d}_b \sigma_{\mu\nu} \gamma_5 u_b
\\ &\oplus& \{ s \leftrightarrow u \} \oplus \{ \bar b \leftrightarrow \bar d \} \oplus \{ s\bar b \leftrightarrow u \bar d \} \, .
\end{eqnarray}
This suggests that if the $X(5568)$ is a $su\bar b\bar d$ tetraquark state of $J^P=1^+$,
its kinematically allowed decay channel would be only the $S$-wave $B_s^* \pi^+$.
This channel was also suggested by the D\O\,Collaboration in the
case that the low-energy photon was not detected~\cite{D0:2016mwd}.
Besides this, its neutral partner may decay into $B_s^0 \gamma$.

\end{enumerate}
We notice that the radiative decay of the neutral partner of the $X(5568)$,
into $B_s^0 \gamma$ and $B_s^* \gamma$, can be used to determine its spin-parity
quantum numbers.

{\it{The prediction of the charmed partner state of the $X(5568)$}}.---The charmed partner state of the $X(5568)$, with the fully open-flavor quark content $su\bar c\bar d$, can be searched
for in many current experiments, such as the Belle, BESIII, LHCb, etc. We can simply replace the bottom quark $b$ by the charm quark $c$
in the interpolating currents $J_3(x)$ with $J^P=0^+$ and $J_{3\mu}(x)$ with $J^P=1^+$, and use them to investigate these
charmed partner states employing the previously obtained formalism.
Similar to $m_b$, we use the ``running mass" of the charm quark in the $\overline{\rm MS}$ scheme, as shown in
Eq. \eqref{parameters}.
If the $X(5568)$ is a $su\bar b\bar d$ tetraquark state of $J^P=0^+$, its charmed partner state is predicted
to have the mass
\begin{eqnarray}
m_{X_c,\, 0^+}=2.55\pm0.09 \text{ GeV}\, ,
\end{eqnarray}
and its possible decay patterns are $D_s \pi$ and $DK$, etc.
While, if the $X(5568)$ is of $J^P=1^+$, its charmed partner state has the mass
\begin{eqnarray}
m_{X_c,\, 1^+}=2.55\pm0.10 \text{ GeV}\, ,
\end{eqnarray}
and its possible decay patterns are $D_s^* \pi$ and $D^*K$, etc.

{\it{Conclusion.}}---In conclusion, we have investigated the recent reported $X(5568)$ resonance as a fully open-flavor
$su\bar b\bar d$ tetraquark state with $J^P=0^+/1^+$. In the framework of QCD sum rules,
we use the scalar $J_3(x)$ and axial-vector $J_{3\mu}(x)$ tetraquark currents of the color anti-symmetric $[\mathbf{\bar 3_c}]_{su} \otimes [\mathbf{3_c}]_{\bar{b}\bar d}$
to perform QCD sum rule analyses, and extract hadron masses to be
$5.58\pm0.14$ GeV and $5.59\pm0.15$ GeV, respectively. These two values are both
in good agreement with the experimental mass of the $X(5568)$. Hence, our results suggest that
the $X(5568)$ can be interpreted as either a scalar or an axial-vector $su\bar b\bar d$ tetraquark state.
We also investigate its possible decay patterns, and propose to observe the radiative decay of the neutral partner of the $X(5568)$
into $B_s^0 \gamma$ and $B_s^* \gamma$ to determine its spin-parity quantum numbers.
Moreover, we predict the mass of the possible charmed partner state of the $X(5568)$, if it exists,
to be around $2.55$ GeV for both cases of $J^P=0^+$ and $1^+$.

\noindent{\bf ACKNOWLEDGMENTS}:
This project is supported by the Natural Sciences and Engineering Research Council of
Canada (NSERC) and the National Natural Science Foundation of China under Grants 11205011,
No. 11375024, No. 11222547, No. 11175073, and No. 11575008; the Ministry of
Education of China (the Fundamental Research
Funds for the Central Universities), 973 program. Xiang Liu is also supported by the National
Youth Top-notch Talent Support Program (``Thousandsof-Talents
Scheme").




\begin{thebibliography}{99}

\bibitem{Agashe:2014kda}
  K.~A.~Olive {\it et al.}  [Particle Data Group Collaboration],
  Chin.\ Phys.\ C {\bf 38}, 090001 (2014).

\bibitem{Liu:2013waa}
  X.~Liu,
  Chin.\ Sci.\ Bull.\  {\bf 59}, 3815 (2014).

\bibitem{Chen:2016qju}
  H.~X.~Chen, W.~Chen, X.~Liu and S.~L.~Zhu, Phys. Rept. 639 (2016) 1-121,
  arXiv:1601.02092 [hep-ph].

\bibitem{Aaij:2015tga}
  R.~Aaij {\it et al.} [LHCb Collaboration],
  Phys.\ Rev.\ Lett.\  {\bf 115}, 072001 (2015).

\bibitem{D0:2016mwd}
  V.~M.~Abazov {\it et al.} [D0 Collaboration],
  arXiv:1602.07588 [hep-ex].

\bibitem{Evdokimov:2004iy}
  A.~V.~Evdokimov {\it et al.} [SELEX Collaboration],
  Phys.\ Rev.\ Lett.\  {\bf 93}, 242001 (2004).

\bibitem{Liu:2004kd}
  Y.~R.~Liu, S.~L.~Zhu, Y.~B.~Dai and C.~Liu,
  Phys.\ Rev.\ D {\bf 70}, 094009 (2004).

\bibitem{Kolomeitsev:2003ac}
  E.~E.~Kolomeitsev and M.~F.~M.~Lutz,
  Phys.\ Lett.\ B {\bf 582}, 39 (2004).

\bibitem{Aubert:2003fg}
  B.~Aubert {\it et al.} [BaBar Collaboration],
  Phys.\ Rev.\ Lett.\  {\bf 90}, 242001 (2003).

\bibitem{Agaev:2016mjb}
  S.~S.~Agaev, K.~Azizi and H.~Sundu,
  arXiv:1602.08642 [hep-ph].

\bibitem{Maiani:2004vq}
  L.~Maiani, F.~Piccinini, A.~D.~Polosa and V.~Riquer,
  Phys.\ Rev.\ D {\bf 71}, 014028 (2005).

\bibitem{Terasaki:2003qa}
  K.~Terasaki,
  Phys.\ Rev.\ D {\bf 68}, 011501 (2003).

\bibitem{Wang:2006uba}
  Z.~G.~Wang and S.~L.~Wan,
  Nucl.\ Phys.\ A {\bf 778}, 22 (2006).

\bibitem{Vijande:2006hj}
  J.~Vijande, F.~Fernandez and A.~Valcarce,
  Phys.\ Rev.\ D {\bf 73}, 034002 (2006)
  [Phys.\ Rev.\ D {\bf 74}, 059903 (2006)].


\bibitem{Jaffe:2004ph}
  R.~L.~Jaffe,
  Phys.\ Rept.\  {\bf 409}, 1 (2005).

\bibitem{Chen:2010ze}
  W.~Chen and S.~L.~Zhu,
  Phys.\ Rev.\ D {\bf 83}, 034010 (2011).

\bibitem{Kleiv:2013dta}
  R.~T.~Kleiv, T.~G.~Steele, A.~Zhang and I.~Blokland,
  Phys.\ Rev.\ D {\bf 87}, 125018 (2013).

\bibitem{Chen:2013aba}
  W.~Chen, T.~G.~Steele and S.~L.~Zhu,
  Phys.\ Rev.\ D {\bf 89}, 054037 (2014).

\bibitem{Du:2012wp}
  M.~L.~Du, W.~Chen, X.~L.~Chen and S.~L.~Zhu,
  Phys.\ Rev.\ D {\bf 87}, 014003 (2013).

\bibitem{Shifman:1978bx}
  M.~A.~Shifman, A.~I.~Vainshtein and V.~I.~Zakharov,
  Nucl.\ Phys.\ B {\bf 147}, 385 (1979).

\bibitem{Reinders:1984sr}
  L.~J.~Reinders, H.~Rubinstein and S.~Yazaki,
  Phys.\ Rept.\  {\bf 127}, 1 (1985).

\bibitem{colangelo}
  P.~Colangelo and A.~Khodjamirian, {\it ``At the Frontier of
  Particle Physics/Handbook of QCD''} (World Scientific,
  Singapore, 2001), Volume 3, 1495.

\bibitem{Nielsen:2009uh}
  M.~Nielsen, F.~S.~Navarra and S.~H.~Lee,
  Phys.\ Rept.\  {\bf 497}, 41 (2010)

\bibitem{Narison:2002pw}
  S.~Narison,
  ``QCD as a theory of hadrons from partons to confinement,''
  Camb. Monogr. Part. Phys. Nucl. Phys. Cosmol. 17 (2002) 1 [hep-ph/0205006]


\bibitem{Narison:2011rn}
  S.~Narison,
  Phys.\ Lett.\ B {\bf 707}, 259 (2012).

\bibitem{Narison:2010cg}
  S.~Narison,
  Phys.\ Lett.\ B {\bf 693}, 559 (2010)
  [Phys.\ Lett.\  {\bf 705}, 544 (2011)].

\bibitem{Kuhn:2007vp}
  J.~H.~Kuhn, M.~Steinhauser and C.~Sturm,
  Nucl.\ Phys.\ B {\bf 778}, 192 (2007).

\bibitem{Chen:2015fwa}
  H.~X.~Chen, E.~L.~Cui, W.~Chen, T.~G.~Steele, X.~Liu and S.~L.~Zhu,
  Phys.\ Rev.\ D {\bf 91}, 094022 (2015).

\bibitem{Chen:2006hy}
  H.~X.~Chen, A.~Hosaka and S.~L.~Zhu,
  Phys.\ Rev.\ D {\bf 74}, 054001 (2006).


\end{thebibliography}
\end{document}